\documentclass[12pt,preprint]{aastex}

\usepackage{graphics}

\shorttitle{PAHs and crystalline silicates in the post-AGB star 
IRAS\,16279$-$4757}
\shortauthors{Matsuura et al.}

\begin{document}

\title{PAHs and crystalline silicates in the bipolar post-AGB star IRAS
16279$-$4757
\thanks{
Based on observations with the European Southern Observatory,
3.6-meter telescope with TIMMI-2 at La Silla.
The proposal number is 71.D-0049.}
\thanks{Based on observations with ISO, an ESA project with instruments
funded by ESA Member States (especially the PI countries: France,
Germany, the Netherlands and the United Kingdom) 
with the participation of ISAS and NASA.}
}

\author{M. Matsuura\altaffilmark{1}, 
           A.A. Zijlstra\altaffilmark{1}, 
           F.J. Molster\altaffilmark{2}, 
           S. Hony\altaffilmark{2}, 
           L.B.F.M. Waters\altaffilmark{3,4},
           F. Kemper\altaffilmark{5,6},
           J.E. Bowey\altaffilmark{7},
           H. Chihara\altaffilmark{8,9}
           C. Koike\altaffilmark{8},
          L.P. Keller\altaffilmark{10}
} 

\altaffiltext{1}{Department of Physics, UMIST, P.O. Box 88, Manchester M60 1QD, UK}
\altaffiltext{2}{ESTEC/ESA, PO Box 299, 2200 AG Noordwijk, The Netherlands}
\altaffiltext{3}{Astronomical Institute `Anton Pannekoek', University of Amsterdam,
       Kruislaan 403, 1098 SJ, Amsterdam,
       The Netherlands}
\altaffiltext{4}{Instituut voor Sterrenkunde, Katholieke Universiteit Leuven, 
        Celestijnenlaan 200B, 3001 Heverlee, Belgium}
\altaffiltext{5}{Department of Physics and Astronomy, University of California, Los
       Angeles, CA 90095-1562, USA}
\altaffiltext{6}{SIRTF fellow}
\altaffiltext{7}{Department of Physics and Astronomy, University College London,
       Gower Street, London WC1E 6BT, UK}
\altaffiltext{8}{Kyoto Pharmaceutical University, Yamashina, Kyoto 607-8412, Japan}
\altaffiltext{9}{Department of Earth and Space Science, Osaka University, Toyonaka, Osaka 560-0043, Japan}
\altaffiltext{10}{Mail Code SR, NASA Johnson Space Center, Houston, Texas 77058, USA}

\begin{abstract}
IRAS\,16279$-$4757 belongs to a group of post-AGB stars showing both
PAH bands and crystalline silicates.  We present mid-infrared images,
that resolve the object for the first time.  The morphology is similar
to that of the `Red Rectangle' (HD\,44179), the prototype object with
PAHs and crystalline silicates.  A two-component model and images
suggest a dense oxygen-rich torus, an inner, low-density carbon-rich
region and a carbon-rich bipolar outflow.  The PAH bands are enhanced
at the outflow, while the continuum emission is concentrated towards
the center.  Our findings support the suggestion that mixed chemistry
and morphology are closely related.  We discuss the ISO/SWS spectra of
IRAS\,16279$-$4757.  Several bands in the ISO/SWS spectrum show a
match with anorthite: this would be the first detection of this
mineral outside the solar system.  Compared to HD\,44179, the shapes
of PAH bands are closer to those of planetary nebulae, possibly
related to a population of small PAHs present HD\,44179, but absent
around IRAS\,16279$-$4757.  Detailed examination of the spectra shows
the individual character of these two objects.  The comparison
suggests that the torus found in IRAS\,16279$-$4757 may have formed
more recently than that in HD\,44179.
\end{abstract}



%
\section{Introduction}

Some post-AGB stars show both PAH and crystalline silicate bands in
their infrared spectra.  The formation history of this mixed chemistry
(oxygen-rich silicates versus carbon-rich PAHs) is not well
understood.  A possibility, but implausible, is that these stars all
evolved from oxygen-rich to carbon-rich within the last few hundred
years \citep{Zijlstra91}.  \citet{Waters98} and \citet{Molster99}
propose that the silicate dust is stored in a long-lived circumbinary
disk.  In this scenario, the PAHs form during a later mass-loss phase,
after the star became carbon-rich, while the gas stored in the disk
retains the chemistry of the earlier, oxygen-rich phase.  Part of the
amorphous silicate dust crystallizes in the disk.  The scenario
explains why there are relatively few post-AGB stars with this mixed
chemistry, and why the silicates have a lower temperature than the
carbon rich dust. It requires all such stars to be binaries, as is the
case for the prototype of the class, the Red Rectangle (HD\,44179; \citet{Waelkens96}).

 IRAS\,16279$-$4757 (hereafter IRAS\,16279) is classified as a
post-AGB star, based on a double peak in its spectral energy
distribution indicating a detached envelope \citep{vanderVeen89}.
Optical spectra suggest a spectral type of G5 \citep{Hu93}.  The
ISO/SWS spectra show crystalline silicates beyond 20\,$\mu$m
\citep{Molster99} and PAH bands are seen in the near-infrared
\citep{vanderVeen89} and ISO spectra.  IRAS\,16279 is therefore a
member of the group of mixed-chemistry post-AGB stars.

 We present TIMMI-2 mid-infrared imaging and spectroscopic data,
resolving IRAS\,16279 for the first time.  In this paper we discuss
the spatial distribution of the different dust components based on
these images, and compare these with predictions from the circumbinary
disk scenario.  We compare the spectra of this object with those of
the prototype mixed-chemistry object, the Red Rectangle.

\section{The observations}

\begin{table}
\begin{caption}{
Filters and flux densities. The last column lists the aperture
used for photometry.}
\label{table-filters}
\end{caption}
\begin{flushleft}
\begin{tabular}{l r@{.}l@{--}r@{.}l r@{.}l r@{.}l c }  
\hline\hline
 Filter & \multicolumn{4}{c}{$\lambda_{\rm 50\% cut}$ [$\mu$m]} &
 \multicolumn{2}{c}{Flux [Jy]} & \multicolumn{2}{c}{Error [Jy]} & Apt. [arcsec] \\\hline
N7.9   & 7&42  &  8&11   &  23&4  &  4&8 & 4   \\
N8.9   & 8&29  &  9&07   &  19&5  &  4&4 & 4   \\
N9.8   & 9&10  & 10&02   &  20&0  &  4&5 & 4   \\
N10.4  & 9&80  & 10&82   &  29&0  &  5&4 & 4   \\
N11.9  & 10&99 & 12&19   &  59&8  &  7&7 & 6   \\
Q      & 17&35 & 18&15   &  172&  & 12&  & 6   \\
\hline \\
\end{tabular}
\end{flushleft}
\end{table}

 We obtained 10 and 18\,$\mu$m images with TIMMI-2 \citep{Kaeufl00} at
the ESO 3.6m telescope on La Silla, on the 27th of May 2003.  The
weather was clear with occasional cirrus.  We used 6 narrow N and
Q-band filters (Table \ref{table-filters}).  The diffraction limit is
about 0.7\,arcsec (first null) in the N-band and 1.4\,arcsec at
Q-band.  The optical seeing of 0.8--1.1\,arcsec in FWHM allowed us to
reach the diffraction limit in the images.  The pixel scale is 0.2
arcsec.  The background was subtracted with nodding and chopping with
an offset of 15 arcsec.  The pipeline-reduced data were used for
imaging.  The results from the aperture photometry are given in
Table\,\ref{table-filters}.  The exposure time is about 16\,minutes
per image.

We also obtained 10\,$\mu$m spectra with TIMMI-2, taken through a
1.2$\times$50 arcsec$^2$ slit, oriented North--South.  The slit was
centered 0.5 arcsec west from the brightest N-band position.  The
resolution is about 160 in $\lambda / \Delta \lambda$.  The background
was subtracted with 15 arcsec chop-and-nod.  We used {\it eclipse} and
{\it IDL} for the data reduction.  The source extends approximately
$6\times8$\,arcsec$^2$, and the spectra are spatially resolved.  The pixel
scale of 0.45 arcsec per pixel was binned to 0.9\,arcsec along the
slit to increase the signal-to-noise ratio.  The total exposure time
is 58\,minutes.

The flux calibration, and the response/atmospheric transmission
correction in the spectra was derived using HD\,169916 (K1III). The
spectral template for HD\,169916 is calculated by \citet{Cohen98}.

We also use archival ISO/SWS spectra, obtained on 21 August 1997 with
a resolution of $\lambda/\Delta\lambda=$250--600.  The aperture is 14$\times$27\,arcsec$^2$ or
larger, sufficient to cover the entire source.  However, there are
several flux discrepancies between the detector bands, probably due to
a slight mis-pointing.  We scaled the separate ISO/SWS bands
(12.45--27.29\,$\mu$m, 27.28--28.90\,$\mu$m, and 28.90--45.38\,$\mu$m)
so as to smoothly connect across the bands, and to simultaneously
agree with the IRAS broad-band flux levels.

\section{Results}

\subsection{Mid-infrared images}

\begin{figure*}[ht]
\centering
\resizebox{\hsize}{!}{\includegraphics*[0,0][480,275]{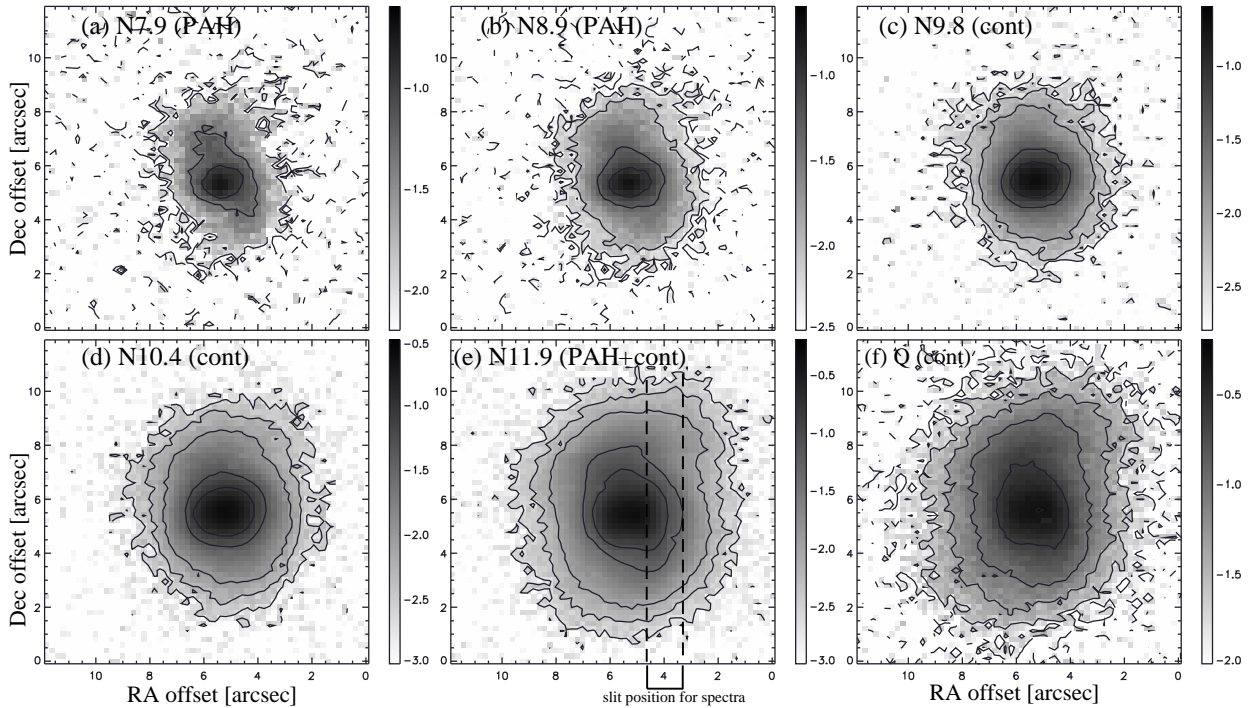}}
\caption{
The images of IRAS 16279$-$4757 in units of $\log F_{\nu}$ in
Jy per pixel.
The object is elongated, especially in the PAH bands.
Contour lines show 0.0025, 0.005, 0.01, 0.05, and 0.1\,Jy per pixel 
for the N-band,
and 0.01, 0.025, 0.05, 0.1, 0.25, and 0.5\,Jy  per pixel for the Q-band.
The dashed lines in (e) show the approximate slit position for the spectra.
}
\label{Fig-images}
\end{figure*}

\begin{figure}[ht]
\centering
\resizebox{\hsize}{!}{\includegraphics*{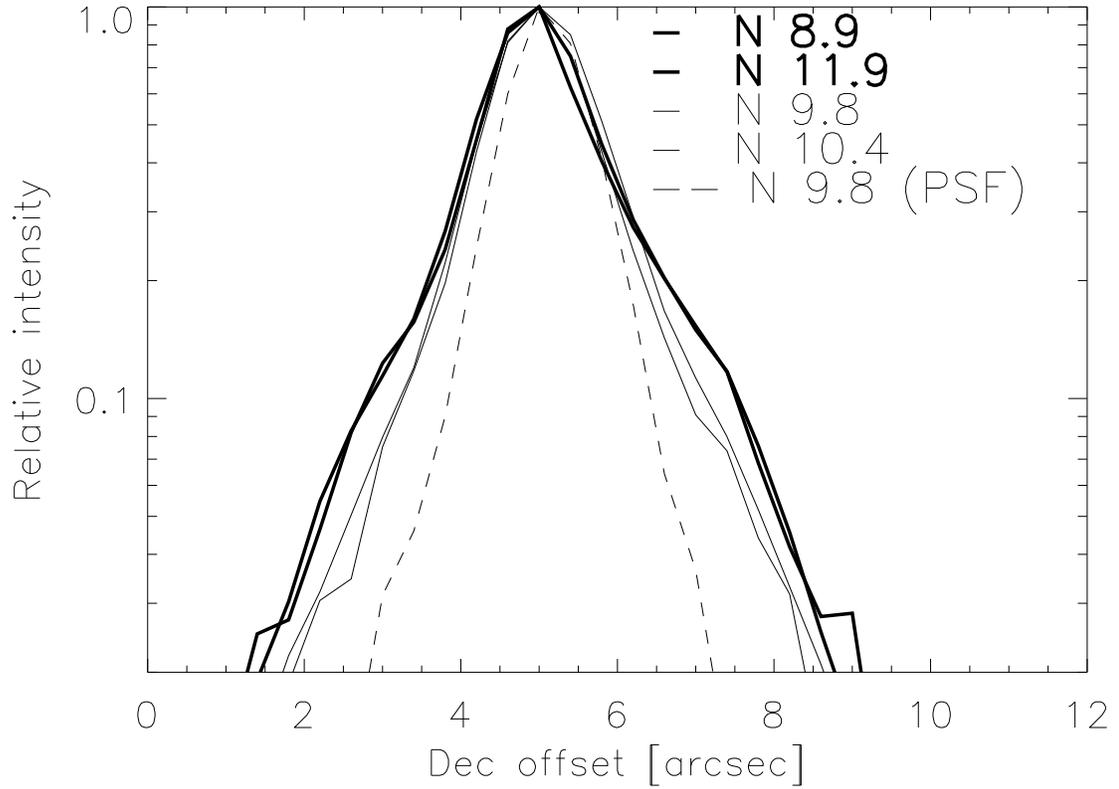}}
\caption{The intensity profiles in declination, across the
source center (corresponding to RA offset = 5.6\,arcsec 
in Fig.\,\ref{Fig-images}).
The peak intensity is normalized to unity.
The profiles of the N8.9 and N11.9 bands, which cover PAH features,
are shown in bold: they have longer tails
than the continuum bands (N9.8 and N10.4; thin lines).
The profile  of the calibration star (HD 169916) at N9.8 band is 
shown as the dashed line.
}
\label{Fig-slice}
\end{figure}

\begin{figure}[ht]
\centering
\resizebox{\hsize}{!}{\includegraphics*{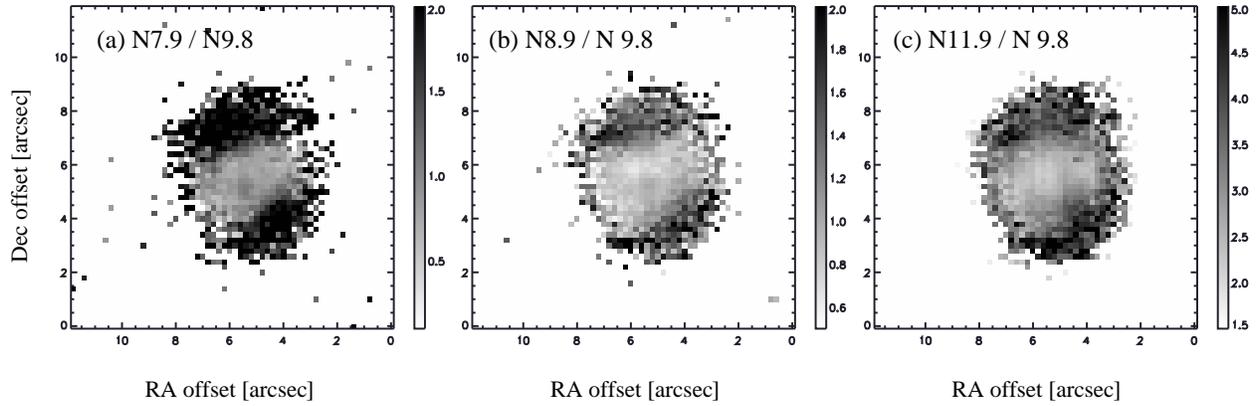}}
\caption{
The intensity ratios of the PAH bands  with respect to the
continuum (N9.8). 
The PAH is enhanced in the outflows.
}
\label{Fig-ratio}
\end{figure}

\begin{figure}[ht]
\centering
\resizebox{\hsize}{!}{\includegraphics*{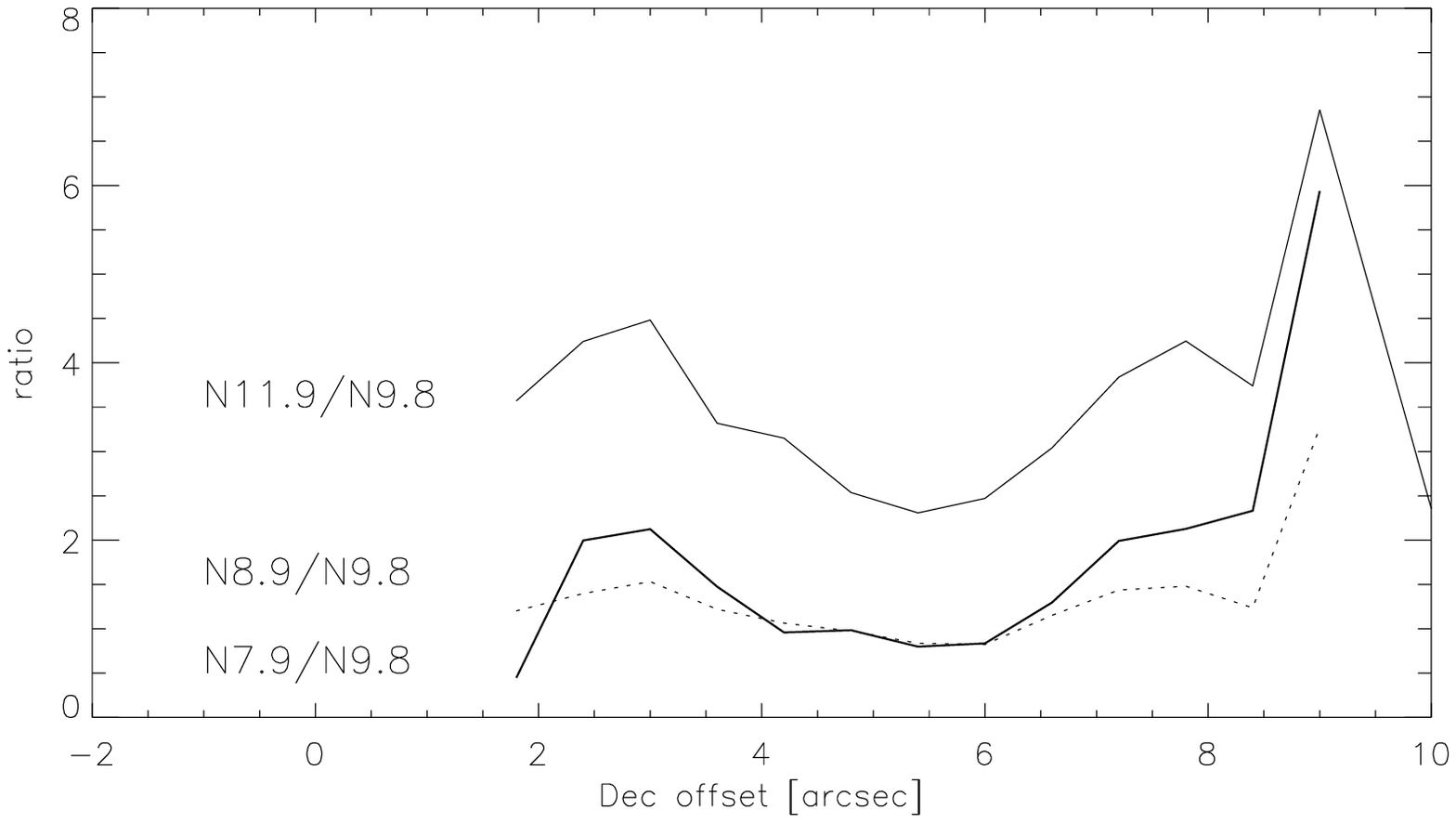}}
\caption{
The intensity ratio of the PAH bands with respect to N9.8
(pseudo-continuum) along declination through the source center 
(RA offset = 5.6\,arcsec). 
The data is smoothed to an 0.4 arcsec grid.
The approximate error levels are about 0.8--1.0.
}
\label{Fig-slice-PAH}
\end{figure}

The TIMMI-2 images of IRAS\,16279 are shown in Fig.\,\ref{Fig-images}.
The object is clearly resolved in all images, and more elongated within
the PAH bands.  This is clearly seen in Fig.\,\ref{Fig-slice} in which
we compare the intensity profiles of the various filters along the
north-south axis with the profiles of a calibration star. To
further quantify the spatial variations in the PAH and continuum
emission, we show the PAH-to-continuum-ratio
(Figs.\,\ref{Fig-ratio} and \ref{Fig-slice-PAH}).  The PAH emission is
enhanced towards the SW and NE, giving the appearance of bipolarity.
At the center, the PAH-to-continuum ratio is low.  The elongated PAH
emission distribution, and the low PAH/continuum ratio at the center
are similar to what is found in the Red Rectangle \citep{Waters96}.

To confirm that the PAH images are more elongated than the continuum
images, we fitted the images with two 2-dimensional Gaussian
functions.  The first Gaussian component represents the bright core,
and the second Gaussian measures the intensity and elongation
direction of the faint, extended emission. The peak positions of the
two Gaussian profiles are fixed at the same place.  The best-fit
parameters are summarized in Table \ref{table-Gauss}.  The PAH bands
(N7.9, N8.9, N11.9) show a relatively stronger second component
(larger $a_2/a_1$ ratio), confirming the extended appearance in these
bands.

The Q-band and N11.9 images (Fig.\,\ref{Fig-images}) show a more
rectangular shape, slightly elongated both along the SW-NE direction
(as found in PAH images) and the SE-NW direction. The rotation angles
change from 160 degrees in the 10-$\mu$m bands, to 75 degrees at
11.9\,$\mu$m, to 10 degrees at 18\,$\mu$m.  The Q-band filter contains
mainly dust continuum (Sec. \ref{model}).  Although the N11.9 band
contains a PAH feature, this contributes  only about 1/2--1/3 of the
intensity, and the rest is continuum.  This rectangular shape might
also be present in other continuum images, however, the emission is
fainter in the shorter-wavelength continuum bands.

\begin{table}
\begin{caption}{
The fitted results of the images with two 2-dimensional Gaussian profiles.
$a$ is the peak intensity in Jy per pixel.
$\sigma_{\rm maj}$ and $\sigma_{\rm min}$ are along the major 
and minor axes in arcsec, respectively.
$\theta$ is the rotation angle of the major axis with respect to North
(clockwise).
The subscripts 1 and 2 show each Gaussian component
($a_1>a_2$).
The first component measures the bright core region, and the
second component shows the more extended emission.
The first component in N7.9 is strongly affected
by the extended component, and $\theta_1$ is totally different
from those in other bands.
}
\label{table-Gauss}
\end{caption}
\begin{flushleft}
\begin{tabular}{l r@{.}l r@{.}l r@{.}l r@{.}l r@{.}l r@{.}l r@{.}l
r@{.}l | r@{.}l r@{.}l r@{.}l r@{.}l r@{.}l  r@{.}l r@{.}l r@{.}l}  
\hline\hline
 & \multicolumn{2}{c}{$\sigma_{1{\rm maj}}$} 
 & \multicolumn{2}{c}{$\sigma_{1{\rm min}}$} 
 & \multicolumn{2}{c}{$a_1$ }        
 & \multicolumn{2}{c}{$\theta_1$}  
 & \multicolumn{2}{c}{$\sigma_{2{\rm maj}}$} 
 & \multicolumn{2}{c}{$\sigma_{2{\rm min}}$} 
 & \multicolumn{2}{c}{$a_2$}         
 & \multicolumn{2}{c}{$\theta_2$}  
 & \multicolumn{2}{c}{$a_{2}$ / $a_{1}$} \\ 
\hline
N7.9  & 1&28 &  0&84 & 0&122 & 152&7 &  1&79 & 0&84 &  0&034 & 136&9 & 0&279\\
N8.9  & 0&66 &  0&40 & 0&117 &  79&5 &  1&64 & 1&06 &  0&056 & 158&0 & 0&479\\
N9.8  & 0&73 &  0&57 & 0&163 &  82&0 &  1&84 & 1&35 &  0&025 & 166&5 & 0&153\\ 
N10.4 & 0&74 &  0&58 & 0&198 &  81&9 &  1&67 & 1&29 &  0&048 & 167&3 & 0&242\\
N11.9 & 0&69 &  0&52 & 0&322 &  86&0 &  1&64 & 1&17 &  0&135 &  74&5 & 0&419\\ 
Q    & 1&10 &  0&79 & 0&266 & 146&1 &  2&39 & 1&73 &  0&213 &   8&4 & 0&801\\ 
\hline 
\end{tabular}
\end{flushleft}
\end{table}

\subsection{TIMMI-2 Spectra}

\begin{figure}[ht]
\centering
\resizebox{\hsize}{!}{\includegraphics*[78,365][585,703]{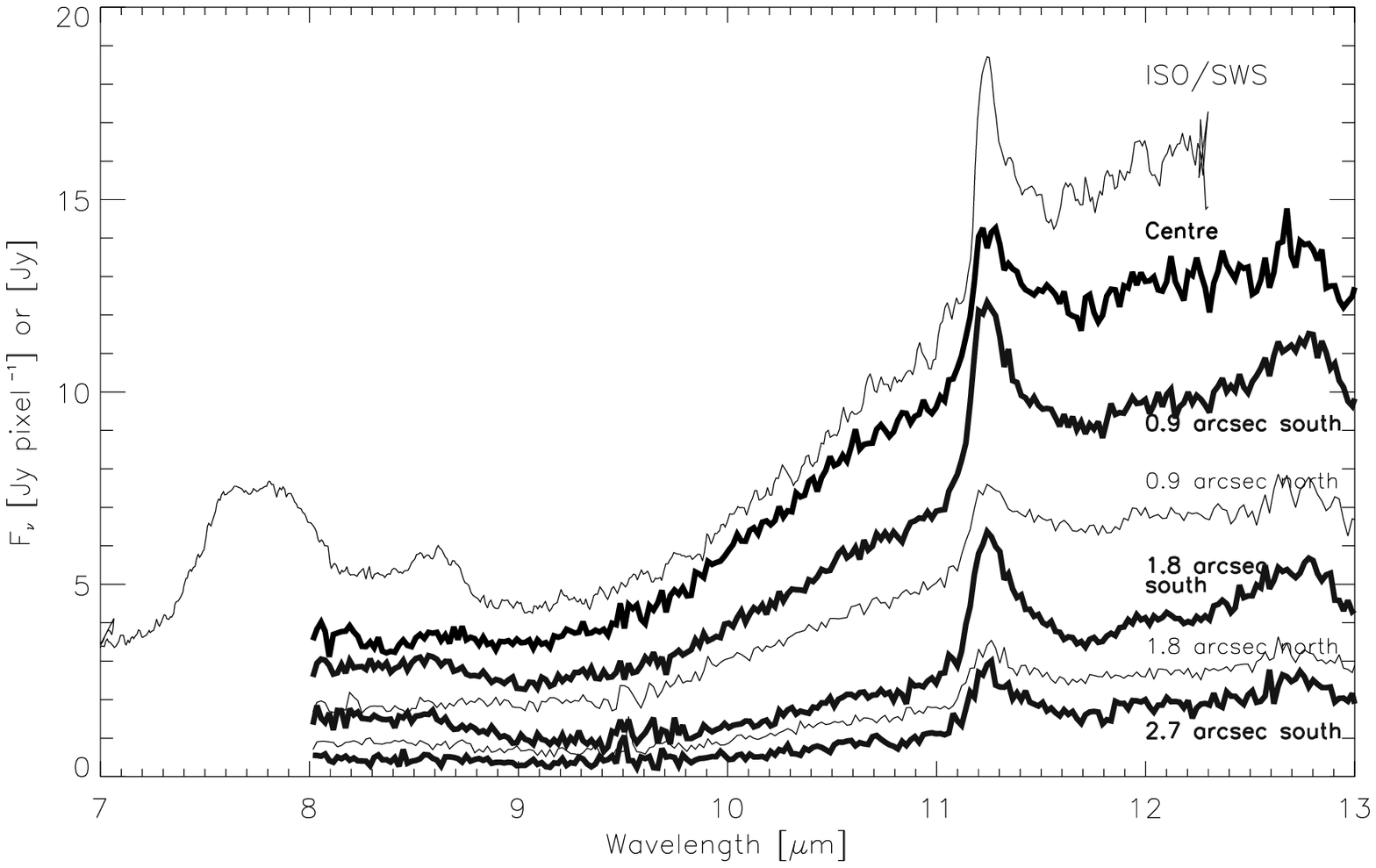}}
\caption{
The N-band spectra with TIMMI-2, compared to ISO/SWS spectra.
The slit position of TIMMI-2 is 
RA offset $\sim$4.0 arcsec, which crosses
the south-west PAH enhanced region.
Flux units of the TIMMI-2 spectra are in Jy per pixel. The ISO/SWS spectra
are in Jy but
are scaled by 1/3.
The 11.2\,$\mu$m PAH feature is strong in the spectra located 0.9\,arcsec 
South,
and changes its shape from north to south.
}
\label{Fig-Nspec}
\end{figure}

The N-band spectra (Fig.\,\ref{Fig-Nspec}) show PAH bands at 7.7, 8.6,
and 11.2\,$\mu$m. The weaker 12.7\,$\mu$m PAH band
\citep*[e.g.][]{Cohen85, Hony01} is also detected in IRAS\,16279.  The
spatially resolved N-band spectra show that the central spectrum
(actually 0.5$^{\prime\prime}$ west of the brightest position) has the
strongest continuum, while the strongest 11.2\,$\mu$m PAH band is
found 0.9$^{\prime\prime}$ south (see also
Fig.\,\ref{Fig-PAHspec}). The spectrum 0.9$^{\prime\prime}$ north
shows only a weak PAH feature. The line-to-continuum ratio is lower at
the center than at 0.9$^{\prime\prime}$ south, in agreement with the
ratio images displayed in Fig.\,\ref{Fig-ratio}.

The lower PAH-to-continuum ratio at the center (Figs.\,\ref{Fig-ratio}
and \ref{Fig-slice-PAH}) is due to the fact that the continuum
intensity declines sharper than the PAH intensity, along the SW-NE
direction (Fig.\,\ref{Fig-PAHspec}).  The PAH intensity itself is
still reasonably high near the center as seen in spectra.

\subsection{Modeling the ISO/SWS spectra}\label{model}

\begin{figure}[ht]
\centering
\resizebox{\hsize}{!}{\includegraphics*{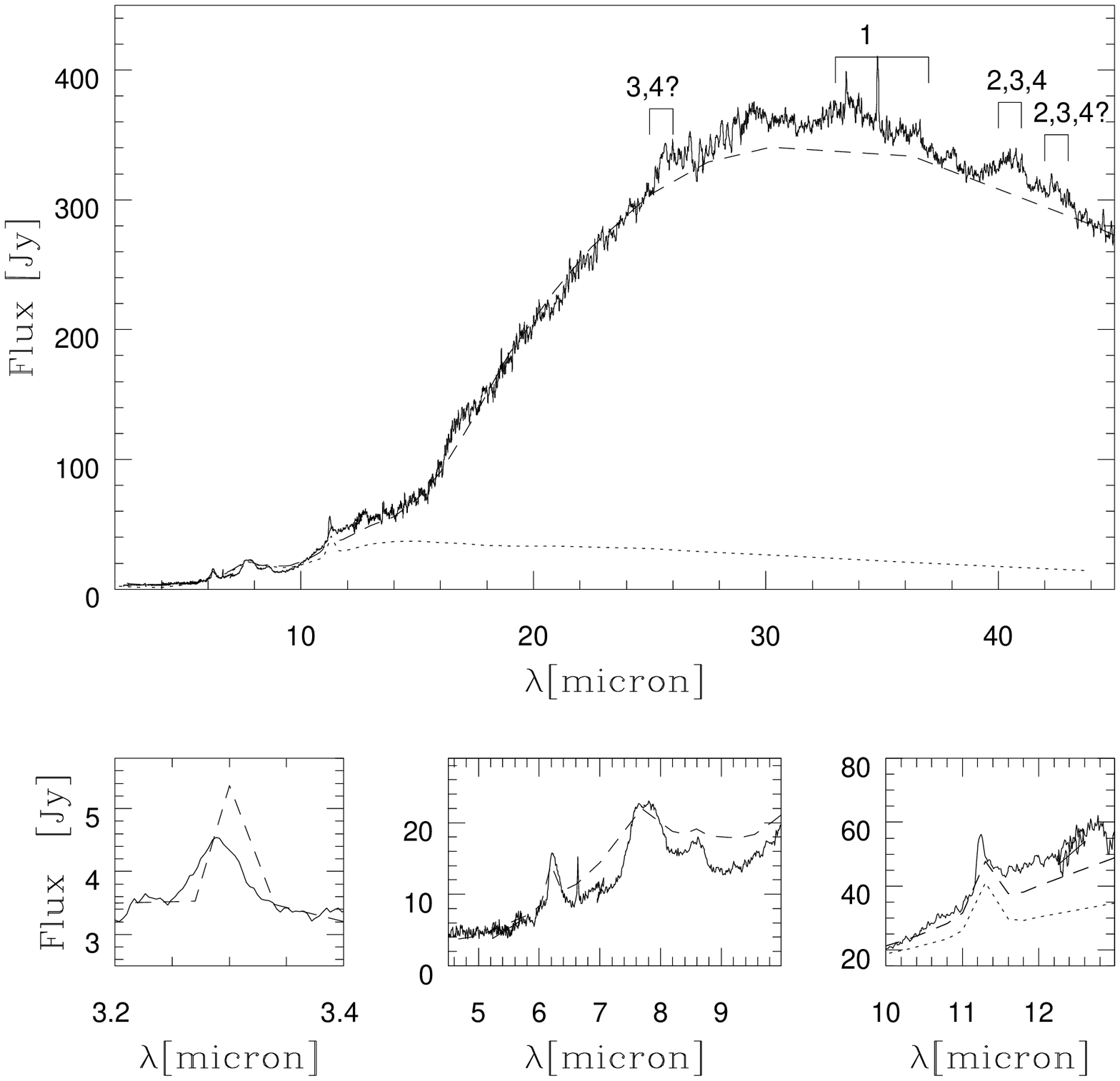}}
\caption{ ISO/SWS spectra. The numbers show the identifications of
crystalline silicate bands:`1' forsterite + plateau, `2' enstatite,
`3' diopside, and `4' anorthite.  Emission lines at 6.63 and
34.8\,$\mu$m could be due to CI.  The dashed line is the two-component
model described in the text. The dotted line represents the continuum
component of the model spectra due to the carbon grains only.  The
spectra of PAH features are enlarged in the bottom.  }
\label{Fig-ISOspec}
\end{figure}

In order to obtain a globally consistent physical picture of the 
source we model the ISO/SWS spectrum. We show the full 2.3--45\,$\mu$m 
spectrum in Fig.\,\ref{Fig-ISOspec}. 
Most of the detected energy falls in the infrared 
range and thus originates from the circumstellar dust. The peak 
wavelength of the energy distribution ($\sim$30\,$\mu$m) implies a 
typical dust temperature of about 100\,K. Below 15\,$\mu$m the spectrum 
is dominate by C-rich material, while beyond 20\,$\mu$m the solid-state 
features can be attributed to crystalline silicates. 

We use a spherically symmetric model consisting of two distinct $n
\propto r^{-2}$ regions, one containing carbon dust and PAHs, and the
other containing silicate dust. The model is described in
\citet*{Siebenmorgen92}, and \citet*{Siebenmorgen94}: it is a 1-d
model which includes scattering and transient heating.  The radiative
transfer is solved independently for the carbon-rich and the
oxygen-rich region. We do not include the possible effect of the
fairly low extinction within the carbon-rich region on the heating of
the (outer) oxygen-rich region.

The distance and extinction are derived assuming a luminosity of
$1\times10^4$\,$L_{\small{\sun}}$. The integrated flux (optical to
infrared photometry from \citet{Hu93}, \citet{VandeSteene00}, IRAS,
and ISO) gives a distance estimate of 2.0\,kpc.  IRAS\,16279 is in the
Galactic plane ($b=+0\fdg09$) and will suffer interstellar extinction.
The optical photometry is reported as $R=18.42$ and $I=14.57$\,mag
\citep{Hu93}.  For a spectral type of G5 (\citeauthor{Hu93}) we derive
$A_V = 14.8$\,mag.  Six field stars within 30\,arcmin show an
interstellar extinction of 1.0\,mag kpc$^{-1}$ out to 1.3\,kpc.  This
suggests an interstellar extinction of $A_V\sim$2.0\,mag.  Most of the
large extinction of IRAS\,16279 is likely circumstellar origin.
Therefore, we ignore the interstellar extinction in the model fit.

Fig.\,\ref{Fig-ISOspec} shows the fitted result.  Up to 10\,$\mu$m,
the emission from carbon material is the dominant source of continuum
emission.  The oxygen shell begins to contribute at 11\,$\mu$m and is
dominant beyond 15\,$\mu$m.  The derived parameters
(Table\,\ref{model.dat}) show that most of the dust is actually cold
oxygen-rich dust: the carbon region is much less dense and contributes
little to the mass.

Table\,\ref{model.dat} shows that the carbon-rich gas is mostly
located inside the oxygen-rich region.  Both regions have a covering
factor (sky coverage as seen form the central star) of about
30--35\,\%. The precise value depends on the choice of parameters, but
all fits we obtained required that the output infrared flux be scaled
down by at least half.  The derived covering factors and the bipolar
morphology in the images suggest that the different dust components
may be located in different directions from the star. The model is
consistent with an oxygen-rich torus, where we assume that the radial
density distribution is the same everywhere within the torus.

The major PAH bands can be reasonably fitted with the model, except
for the one at 8.6\,$\mu$m (this problem is also found by
\citet{Siebenmorgen94}).  The oxygen shell gives less than half the
observed $A_V$ and a weaker-than-observed amorphous silicate
absorption feature (seen against the carbon-continuum): the remainder
may be due to interstellar absorption, or may be due to geometrical
effects. The 10\,$\mu$m absorption is seen  mainly against the inner arcsecond,
carbon-rich dust.  For a $r^{-2}$ shell, the tangential line of sight
can have a much higher extinction, and depending on the viewing angle,
this can affect the carbon-rich region.  The 1-d nature of the model
means that self-absorption within the dust regions is not well
described. The line of sight to the star also shows higher extinction,
but it is clear that a global 1-d model may well underestimate  the column
density in individual lines of sight.

The carbon rich shell has an outer radius of $5\times 10^{16}$\,cm
(1.8\,arcsec) in the model, which is close to $\sigma_{\rm 2 maj}$
measured in the PAH bands.  The inner radius of carbon-rich shell is
$1\times10^{16}$\,cm (0.3\,arcsec). The inner radius is not
clearly seen in the image because the spatial resolution is not high
enough and because the point spread functions of the central star and
the shell are merged.

The model fitting yields several constraints: 
\begin{itemize}
\item[(1)] To excite the PAH, the model requires
a stellar temperature of 6\,000\,K or more; otherwise the
bands disappear due to lack of exiting photons.
This includes absorption by neutral PAHs. 
The PAH emission features are thought to arise from
mixture of neutral and charged PAHs.
PAH ions can absorb
in the optical  \citep[c.f.][]{Li02} and this would allow a lower
stellar temperature. \citet{Li02} show that weak PAH bands can be seen
for stellar temperatures as low as 3000\,K, but reach full strength only
above 6000\,K. 
Additionally, the presence of 11.2\,$\mu$m feature, which attributes to neutral
PAHs \citep{Hony01} suggests that UV radiation is necessary and
extremely low temperature is not possible.
The observed PAH bands are compatible with a G-type
central star.
\item[(2)] The PAH bands are only seen
in relatively unshielded (low extinction) gas.  Within the model, only
a low density shell contributes to the PAH bands.  Higher density
carbon regions, if present, would not contribute to the observed PAH
bands: the excitation of the PAHs would be suppressed by internal
extinction.  We therefore have few constraints on higher density
carbon-rich gas.
\item[(3)] The oxygen-rich gas is located relatively far out
(inner radius of 1\,arcsec) and has much higher mass
than the carbon-rich gas.  The low temperatures of this gas are
consistent with the observed crystalline silicate bands (70--110\,K;
Sect.\,\ref{sect-crystalline}).
\end{itemize}

\begin{table}
\caption[]{\label{model.dat} Model parameters of the two-shell model.
  }
\begin{flushleft}
\begin{tabular}{c c c c }  
\hline\hline
&   & C shell     & O shell  \\\hline
 $r_{\rm in}$    & [cm]  & $1\times 10^{16}$ & $4\times 10^{16}$ \\
 $r_{\rm out}$   & [cm]  & $5\times 10^{16}$ & $8\times 10^{16}$ \\
 $M_{\rm shell}$ & [M$_\odot$] & $3 \times 10^{-3}$ & 2.1 \\
 $A_{V}$         & [mag] & 0.6 & 6.4 \\
 $T_{\rm d}$     & [K]   &  330--140 & 125--57 \\
Coverage         & [\%]  & 30 & 35 \\
\hline \\
\end{tabular}
\end{flushleft}
\end{table}

\section{Dust features}

The N-band images clearly show that the PAH emission originates from a
bipolar geometry. This further strengthens the link between this
source and HD\,44179, and the link between the mixed chemistry nature and
a bipolar (+disk) geometry. It is interesting to further explore the
comparison between IRAS\,16279 and the Red Rectangle as the prototype
of the dual chemistry sources. In the following we will compare the
PAH emission features between the two sources, and 
we present a detailed comparison of the
crystalline silicate bands.

\begin{figure}[ht]
\centering
\resizebox{\hsize}{!}{\includegraphics*[75,365][465,855]{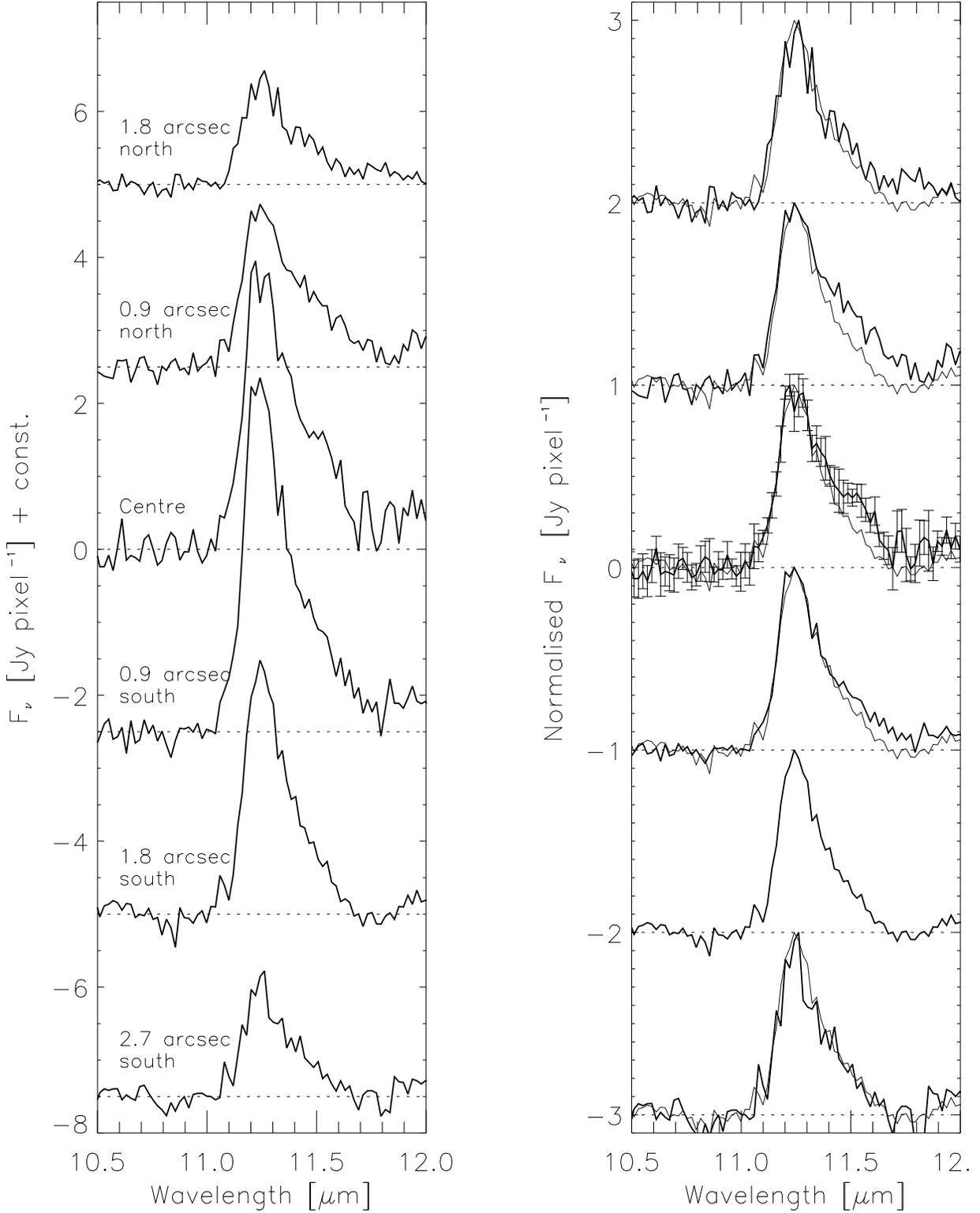}}
\caption{ The 11.2\,$\mu$m PAH feature after subtraction of the continuum
(left). The continuum is estimated using a linear fit between 10.5--11.0 and
11.9--12.4\,$\mu$m.  The PAH feature is strongest at the SW outflow
(0.9$^{\prime\prime}$ south). (Right) To examine the variation in the
shape, the individual spectra are normalized to the peak
intensity. We plot the spectra at 1.8$^{\prime\prime}$ South in gray
lines as references.  Near the center, the feature is broader at the
red wing of the 11.2\,$\mu$m feature.  The error at certain wavelengths
is estimated from the standard deviation of the nearby 5 pixels with the same
wavelength in target and calibration data.  }
\label{Fig-PAHspec}
\end{figure}

\subsection{PAH bands}

\subsubsection{Integrated spectrum}

\begin{figure*}[ht]
  \resizebox{\hsize}{!}{\includegraphics*{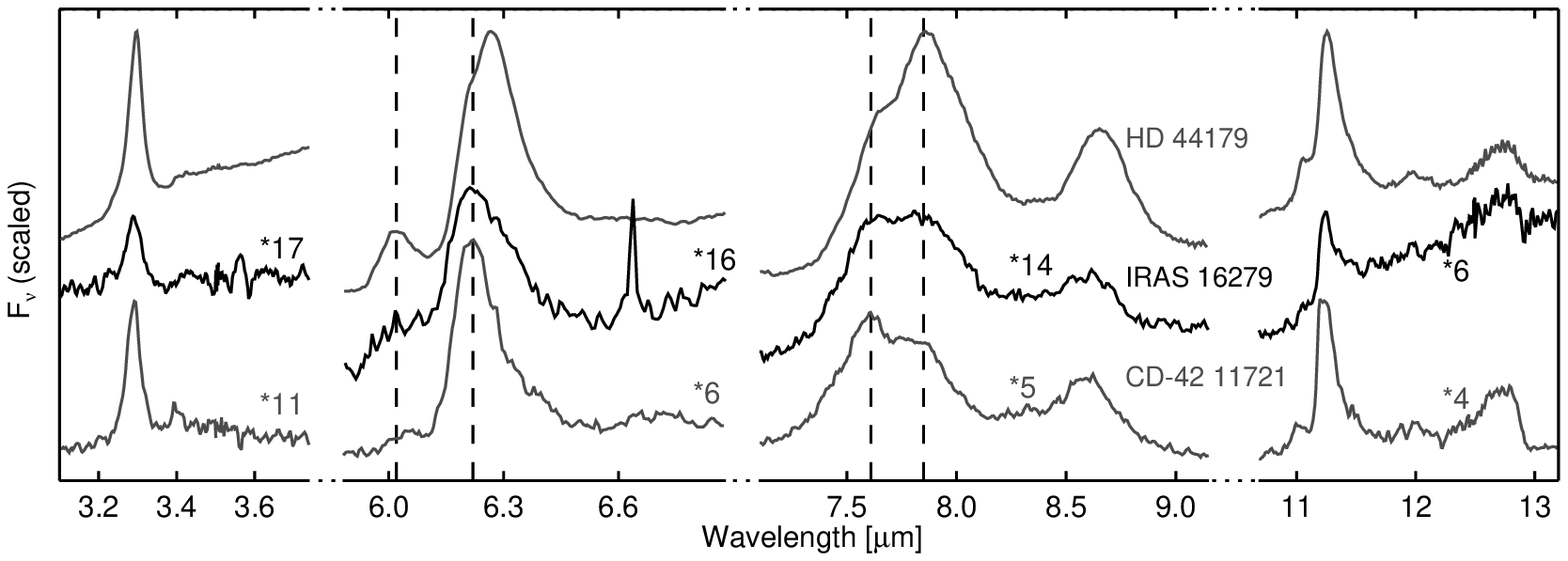}}
  \caption{The comparison between the PAH bands in the ISO/SWS spectra
    of HD\,44179, IRAS\,16279 and the reflection nebulosity near
    CD-42~11721 (top-to-bottom). The dashed lines indicate
    where the IRAS\,16279 spectrum differs most from the ISM PAH
    spectrum (as represented by CD-42\,11721) and HD\,44179.
The spectra are binned to a resolution of $\lambda/\Delta\lambda=200$}
  \label{Fig-PAHs}
\end{figure*}

In Fig.\,\ref{Fig-PAHs} we show the ISO PAH spectra of IRAS\,16279,
HD\,44179 as the prototype of the mixed chemistry post-AGB stars and
CD-42\,11721 as a prototypical ISM/star-forming-region PAH spectrum.
The profile features are very similar in the 3\,$\mu$m and the 10 to
13\,$\mu$m region (C--H modes) while large differences are found in
the PAH spectrum from 6 to 9 $\mu$m (C--C modes). This strong
variation in the 6 to 9\,$\mu$m PAH bands when compared to the other
PAH bands is a common feature of the interstellar PAH spectra
\citep{vanDiedenhoven04}.  The profiles of the 6--9\,$\mu$m PAH bands
of IRAS\,16279 are intermediate between HD\,44179 and CD-42\,11721
\citep{Peeters02}.  The 6.2\,$\mu$m feature of IRAS\,16279 is broader
than in CD-42\,11721.  This broadening is due to an extra contribution
near 6.05\,$\mu$m which is also present in HD\,44179 and a weak
emission band which peaks near 6.28\,$\mu$m.  The latter band is
dominant in the spectrum of HD\,44179 while only marginally detected
in IRAS\,16729, where the usual 6.22\,$\mu$m band is the strongest.

The PAH bands from 7 to 9\,$\mu$m (regularly dubbed the ``7.7
complex'') also attest to the intermediate nature of IRAS\,16279 PAH
emitters. In the ISM and in star-forming regions, usually the
7.6\,$\mu$m band is dominant, while in planetary nebulae and post-AGB
stars the 7.8\,$\mu$m band is strongest \citep{Bregman89, Cohen89,
Peeters02}.  In IRAS\,16279 both bands are roughly equally strong.

\citet{Hony01} found that the band strength ratio of the 11.2 to
12.7\,$\mu$m bands ($I_{11.2}/I_{12.7}$ where $I_{11.2}$ and
$I_{12.7}$ represent the integrated flux of each band) correlates with the
source type. In the planetary nebulae this ratio is very high as is
the case for HD\,44179 ($I_{11.2}/I_{12.7}$=5.0). The band strength
ratio found in IRAS\,16279 ($I_{11.2}/I_{12.7}$=2.0) is not typical
for evolved stars.

When comparing the band strength of the 3.3\,$\mu$m band with the
longer wavelength PAH bands we find that the 3.3\,$\mu$m of
IRAS\,16279 is much weaker than in HD\,44179. The ratios of the
integrated band strengths, $I_{11.2}/I_{3.3}$, are 1.9, 5.2 and 3.9 in
HD\,44179, IRAS\,16279, CD-42\,11721 respectively.  The ratios of
IRAS\,16279 are closer to CD-42\,11721. This may point to a population
of smaller PAHs around HD\,44179 which does not exist around the other
two sources.  Such small PAHs might be easily destroyed when exposed
to the hard UV radiation present in the ISM and planetary nebulae.

\subsubsection{Spatially resolved spectroscopy}

In Fig.\,\ref{Fig-PAHspec} we show the profile of the 11.2\,$\mu$m
band as extracted from the N-band spectra for different positions. On
the right we compare the normalized profiles.  The profile varies with
the position along the slit, being broader at the central position and
0.9$^{\prime\prime}$ north. A longer tail at 11.5\,$\mu$m is usually
interpreted as being due to hotter PAHs \citep*{Tielens99, Pech02}.
Alternatively, a different profile might arise from a variation of the
PAH population as a function of distance to the source. For example,
PAHs with two neighboring C--H bonds may give rise to emission in the
red wing of the 11.2\,$\mu$m solo C--H mode \citep{Hony01}. We prefer
the hot PAH interpretation, because of the fact that the sharp
blue-side rise stays the same in the all spectra, which is difficult
to envisage in the scenario of a strongly varying PAH population.
Furthermore, the most pronounced tail is found closest to the central
star, where multiple photon heating may play a r\^ole and therefore
hotter PAHs might be expected.

\citet{Miyata01} show spatially resolved N band spectra of HD\,44179
in which they also find variations of the emission features as a
function of distance to the star. Like in IRAS\,16279 the 11.2\,$\mu$m
band diminishes close to the central star. However, unlike in
IRAS\,16279, the HD\,44179 spectrum near the central star is dominated
by bands at $\sim$11.0 and $\sim$11.9\,$\mu$m. These bands could be
related to the crystalline enstatite that is present in the disk.  Due
to this possible confusion between the different contributors to the
emission in HD\,44179 it is not feasible to determine the band profile
of the 11.2\,$\mu$m nearest the central star.

Concluding, we can state that there are significant differences in the
PAH spectrum of HD\,44179 and IRAS\,16279, both in band strength ratios
and profile features, where the latter source exhibits an PAH spectrum
intermediate between HD\,44179 and the prototypical ISM PAH spectrum.
In this sense the PAH spectrum of IRAS\,16279 compares better, in bands
strengths and feature profiles, with the young C-rich planetary
nebulae IRAS\,21282+5050 \citep{Hony01, Peeters02}.

\subsection{Crystalline silicate bands}\label{sect-crystalline}

\begin{figure}[ht]
\resizebox{\hsize}{!}{\includegraphics*[80,367][548,702]{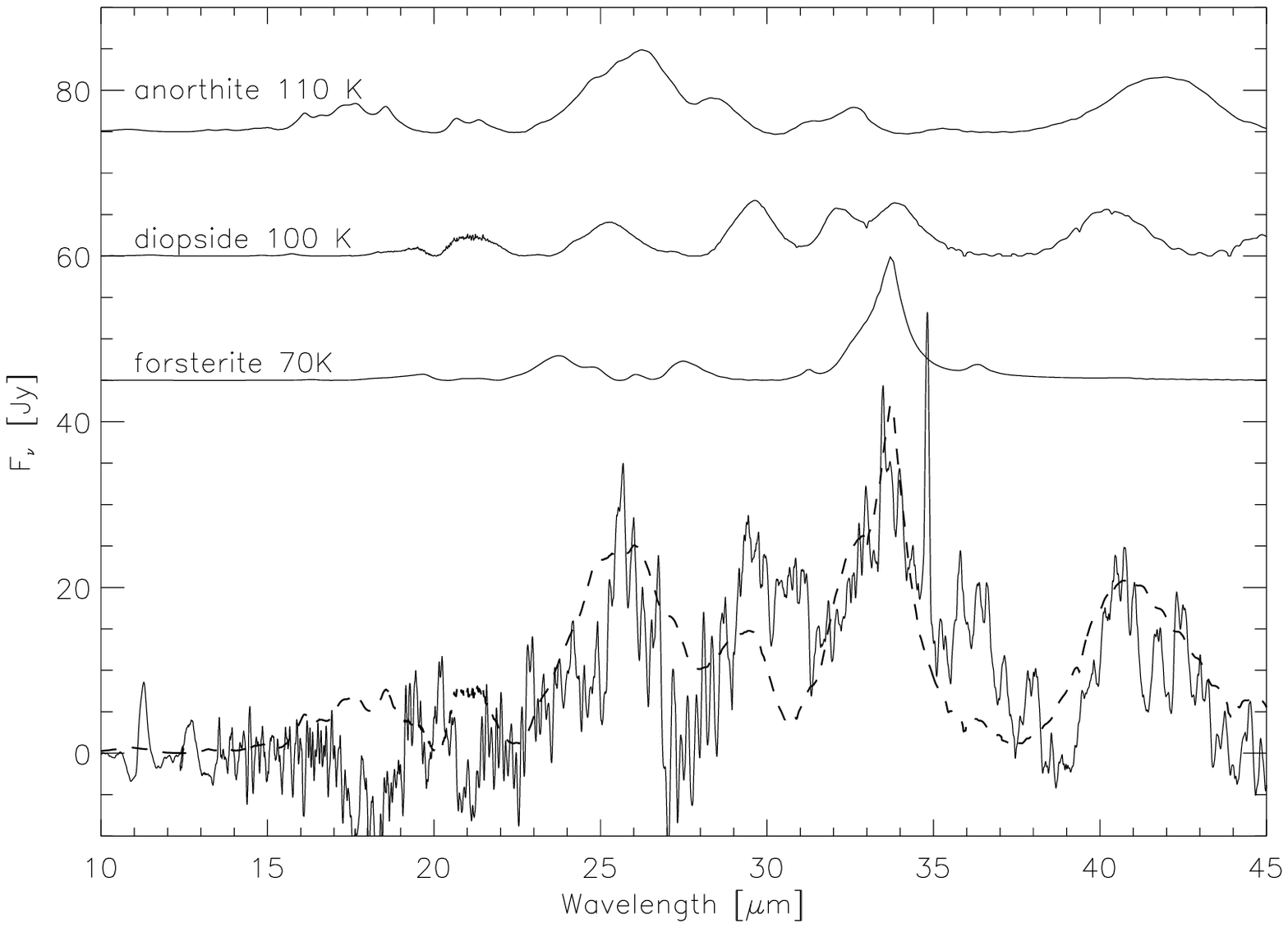}}
  \caption{
The continuum-subtracted ISO/SWS spectra (solid line).
The dashed line shows the fitting of the crystalline bands \citep{Molster02b}.
The dust properties are measured by 
\citet{Koike99, Koike00, Chihara03}.
}
  \label{Fig-cry}
\end{figure}

 The ISO/SWS spectra beyond 20\,$\mu$m show the solid state features
which are attributed to crystalline silicates
(Fig.\,\ref{Fig-ISOspec}).  Continuum subtracted spectra are shown in
Fig.\,\ref{Fig-cry}.  The continuum is obtained with a quadratic fit
(polynomial fit to the three neighboring points) to the smoothed
spectra.

The feature at 33.7\,$\mu$m has been identified with crystalline
olivine (Mg$_{2x}$Fe$_{2-2x}$SiO$_4$, with $x$ between 0 and 1). Its
position fits best with that of the pure Mg-olivine, i.e. forsterite
(Mg$_2$SiO$_4$) \citep{Waters96, Molster02a}.  The band ratio depends
on the temperature \citep[ chap.~1]{Molster00}.  The absence of a
feature around 23.7\,$\mu$m shows that forsterite is below 100\,K.
The plateau up to 37\,$\mu$m, which has no clear mineralogical
identification, is prominent in this star.

The feature at 40.5\,$\mu$m is normally attributed to pyroxene-like
silicates ({\it Me}SiO$_3$, where {\it Me} stands for metal).  The
presence of features also around 43\,$\mu$m is normally an indication
for the presence of enstatite (MgSiO$_3$).  However, the lack of
enstatite features below 30\,$\mu$m makes this identification
questionable.  The ISO/SWS band 3E (27.5--29\,$\mu$m) is known for its
mediocre performance. This could be the reason for the lack of
enstatite features around 28\,$\mu$m. Alternatively, enstatite can
also simply be very cold just like the forsterite.  Whatever the
reason is, it should be noted that the 40.5\,$\mu$m feature is rather
strong with respect to the 43\,$\mu$m feature if it is purely due to
enstatite.  This implies that more materials might be present.

There are some other features present around 26, 29 and 32\,$\mu$m.
These three positions together with the 40.5\,$\mu$m feature could fit
with the features of diopside (CaMgSi$_2$O$_6$) measured in the
laboratory \citep{Koike00}.  The slight wavelength shift around 26 and
40.5\,$\mu$m with respect to the laboratory, and, in addition, the
lack of a clear feature around 21\,$\mu$m makes the identification
less conclusive.  But these differences might be explained by
temperature and grain size and shape effects as is also observed for
other silicates, like forsterite \citep{Bowey02, Molster02a,
Molster02b}.

Another mineral that might explain some of the features is anorthite.
Properties of anorthite have been measured in the laboratory
\citep{Chihara03, Keller03}.  The profile presented in
Fig.\,\ref{Fig-cry} is for a sample of An$_{96}$ in \citet{Chihara03}.
It matches with the feature at 26\,$\mu$m, the long wavelength side of
the 40\,$\mu$m complex and the extra feature just before the
33.7\,$\mu$m forsterite feature.  This would be the first
identification of anorthite outside the solar system.  The lack of
anorthite features in the ISO-spectrum around 18\,$\mu$m might be
explained by low temperature, or interstellar absorption.

We would require future spectra beyond 45\,$\mu$m to confirm the
presence of anorthite and diopside, which have stronger bands in these
wavelength.  We should also stress here that we are dealing with an
artificial continuum and this can have some influence on the strength of
the features.

\section{Discussion}

The mid-infrared images of IRAS\,16279 show an elongation in the PAH
band and a rectangular shape in the N11.9 and Q-band images.  At the
center, the PAH-to-continuum ratio decreases.  There is some
resemblance to the Red Rectangle (HD\,44179) \citep{Waters98}. In the
Red Rectangle, the silicates are thought to be located in a disk and
the PAHs are in the perpendicular polar flows. The images of
IRAS\,16279 strengthen this link between morphology and mixed
chemistry as proposed by \citet{Waters96} and \citet{Molster99}.

The temperature of the crystalline silicate bands in HD\,44179 is
about 135\,K (enstatite; \citet{Molster02b}), while it is below 110\,K
in IRAS\,16279.  Nevertheless, both objects show a lower temperature
in the silicates than in the PAH region.  In fact, all crystalline
silicates in post-AGB stars tend to show low temperatures
(100--250\,K) \citep{Molster02b}.

In the case of IRAS\,16279, the model suggests that the PAH emission
in the SW and NE regions (Fig.\ref{Fig-images}) cannot be shielded by
the dusty oxygen-rich regions. The oxygen-rich material should be
configured as an torus rather than a shell, allows radiation to leak
towards the SW and NE directions.  The model explains the PAH excess
at 2 arcsec: the decreasing temperature suppresses the continuum but
not the PAH bands.

This structure, with a low-density carbon-rich region and an obscured,
dense oxygen-rich torus, agrees with the (Red Rectangle)
oxygen-rich-disk scenario of \citeauthor{Waters96} and
\citeauthor{Molster99} But the oxygen-rich region in IRAS\,16279 is
several times larger than seen in the Red Rectangle.  A circumbinary
disk is expected to be compact in order to store the oxygen-rich gas
over a long time.  It is not clear whether this has happened in
IRAS\,16279.  The extended torus should disrupt faster than the disk
of the Red Rectangle.
Its formation may
have happened relatively recent. 




The crystallization of amorphous silicates occurs via heating and
subsequent cooling of the grain.  This may occur slowly at low
temperature in a long-term stable disk, under the influence of UV
radiation (as in the Red Rectangle), or quickly in the AGB wind at
very high mass-loss rates, through high temperature annealing
\citep{Waters96, Sylvester99}.  Assuming an expansion velocity of
20\,km\,s$^{-1}$, the mass-loss rate of IRAS\,16279 was of order $5
\times 10^{-4}\,\rm M_\odot\,yr^{-1}$, sufficient for high temperature
annealing in the outflow to occur. Therefore, crystallization itself
may occurred already in the AGB outflow.  Part of the crystallized
silicate is stored in the disk later, and part of it may remain in the
outflow.

It is not clear that the torus is stable long enough for long-term
crystallization.  This may be solved by shocks addressed by
\citet*{Harker02} who found a 5\,km\,s$^{-1}$ shock is sufficient
for annealing comets. This velocity range might be possible in AGB or
post-AGB wind interaction with the torus.  In this case, the higher
density of the torus is more likely to obtain a higher rate of
crystallization, and this may compensate the short life of the torus
than the disk.


%

The crystalline silicate bands in IRAS\,16279 are significantly weaker
than in the HD\,44179 \citep{Waters98}.  The weakness of the features
indicate that either they are less abundant than in HD\,44179, or the
temperature difference with the hotter amorphous silicates
(responsible for the continuum) is larger in IRAS\,16279. Although a
difference in abundance can clearly not be excluded, the low
temperature of the crystalline silicates seems to be supported by the
absence of detectable features below 30\,$\mu$m.


\begin{acknowledgements}
 We are grateful for the support from ESO staff members during the
TIMMI-2 observations, especially Drs. Doublier and Comeron.  We
acknowledge Drs. Van de Steene and van Winckel for the discussion of
Br$\gamma$ in this object.  We thank ESA for maintaining the ISO data
archives and ISO/SWS data.  M.M. and J.E.B. are financially supported
by PPARC, and F.K. is supported by NASA through the SIRTF Fellowship
Program, under award 011 808-001.

\end{acknowledgements}


\begin{thebibliography}{}


\bibitem[Bowey et al.(2002)]{Bowey02}
 Bowey, J.E., Barlow, M.J., Molster, F.J., et al.
 2002, MNRAS, 331, L1


\bibitem[Bregman (1989)]{Bregman89}
 Bregman, J. 1989, in IAU Symp. 135, Interstellar Dust, 
 ed. L.J. Allamandola, \& A.G.G.M. Tielens,
 Kluwer Academic Publishers, Dordrecht, p.109


\bibitem[Brucato et al.(1999)]{Brucato99}
Brucato J.R., Colangeli L., Mennella V., Palumbo P. \& Bussoletti
E. A\&A 1999, 348, 1012

\bibitem[Chihara et al.(2003)]{Chihara03}
 Chihara H., Koike, C., Tsuchiyama, A., in preparation


\bibitem[Cohen(1985)Cohen, Tielens, Allamandola]{Cohen85}
 Cohen, M., Tielens, A.G.G.M.,
 Allamandola, L. J.
 1985, ApJ 299, L93

\bibitem[Cohen(1998)]{Cohen98}
 Cohen M. 1998, AJ 115, 2092

\bibitem[Cohen et al.(1989)]{Cohen89}
 Cohen, M., Tielens, A.G.G.M., Bregman, J. et~al., 1989, \apj, 341,
  246






\bibitem[Harker, Desch(2002)Harker, Desch]{Harker02}
 Harker, D.E., Desch, S.J., 
 2002, ApJ 565, L109


\bibitem[Hony et al.(2001)]{Hony01}
 Hony, S., Van Kerckhoven, C.,
 Peeters, E., Tielens, A.G.G.M.,
 Hudgins, D.M., Allamandola, L.J.
 2001, A\&A 370, 1030

\bibitem[Hu et al.(1993)]{Hu93}
 Hu, J.Y., Slijkhuis, S., de Jong, T., Jiang, B.W.
 1993, A\&AS 100, 413

\bibitem[Hu et al.(1994)]{Hu94}
 Hu, J.Y., Te Lintel Hekkert, P., Slijkhuis, F., Baas, F., Sahai, R.,
 Wood, P. R.
 1994, A\&AS 103, 301

\bibitem[Jura et al. (1995)Jura, Balm, Kahane]{Jura95}
 Jura, M., Balm, S.P., Kahane, C.,
 1995, ApJ 453, 721

\bibitem[K\"aufl et al.(2000)]{Kaeufl00}
 K\"aufl, H. U., Ageorges, N., Dietzsch, E., et al. 2000,
 The Messenger, ESO, 102, 4  

\bibitem[Keller et al.(2003)]{Keller03}
 Keller L., in preparation

\bibitem[Koike et al.(1999)]{Koike99}
 Koike, C., Shibai, H., Suto, H. et al.
 1999 in Proceedings of the 32nd ISAS Lunar and Planetary Symposium 32, p175

\bibitem[Koike et al.(2000)]{Koike00}
 Koike, C.; Tsuchiyama, A., Shibai, H., et al.
 2000, A\&A, 363, 1115


\bibitem[Li, Draine(2002)]{Li02}
 Li, A., Draine, B. T.
 2002, ApJ 572, 232

\bibitem[Miyata et al.(2001)]{Miyata01}
 Miyata, T., Kataza, H., Okamoto, Y., et~al., 2001, in Post-AGB Objects
  as a Phase of Stellar Evolution, 
  ed. R. Szczerba \& S. K. G\'orny,
  Astrophysics and Space Science Library, 265, 351

\bibitem[Molster et al.(1999)]{Molster99}
 Molster, F.J., Yamamura, I., Waters, L.B.F.M. et al.
 1999, Nature 401, 563

\bibitem[Molster(2000)]{Molster00}
 Molster, F.J., PhD thesis, 2000, University of Amsterdam

\bibitem[Molster et al.(2002a)Molster, Waters, Tielens]{Molster02a}
 Molster, F.J., Waters, L.B.F.M., Tielens, A.G.G.M.
 2002a, A\&A 382, 222

\bibitem[Molster et al.(2002b)]{Molster02b}
 Molster, F.J., Waters, L.B.F.M.,
 Tielens, A.G.G.M., Koike, C.,
 Chihara, H. 2002b, A\&A 382, 241

\bibitem[Pech et al.(2002)Pech, Joblin, Boissel]{Pech02}
 Pech, C., Joblin, C., Boissel, P.
 2002, A\&A 388, 639

\bibitem[Peeters et al.(2002)]{Peeters02}
 Peeters, E., Hony, S., Van Kerckhoven, C.,
 Tielens, A.G.G.M., Allamandola, L.J.,
 Hudgins, D.M., Bauschlicher, C.W.
 2002, A\&A 390, 1089




\bibitem[Siebenmorgen, Kr\"ugel(1992)]{Siebenmorgen92}
 Siebenmorgen, R., Kr\"ugel, E.
 1992, A\&A 259, 614


\bibitem[Siebenmorgen et al.(1994)Siebenmorgen, Zijlstra, Kr\"ugel]{Siebenmorgen94}
 Siebenmorgen, R., Zijlstra, A.A., Kr\"ugel, E.
 1994, MNRAS 271, 449


\bibitem[Sylvester et al.(1999)]{Sylvester99}
 Sylvester, R.J., Kemper, F.,
 Barlow, M.J., de Jong, T.,
 Waters, L.B.F.M., Tielens, A.G.G.M.,
 Omont, A.
 1999, A\&A 352, 587

\bibitem[Thompson et al.(2002)]{Thompson02}
 Thompson S. P., Fonti S., Verrienti C., Blanco A., Orofino V., \&
 Tang C. C. A\&A 2002, 395, 705


\bibitem[Tielens et al.(1999)]{Tielens99}
 Tielens, A.G.G.M., Hony, S.,
 van Kerckhoven, C., Peeters, E., 1999,
 `The Universe as Seen by ISO', 
 ed. P. Cox \& M. F. Kessler,
 ESA-SP 427, p579

\bibitem[van Diedenhoven et al.(2004)]{vanDiedenhoven04}
 van Diedenhoven et al., 2004, A\&A in preparation

\bibitem[van der Veen et al.(1989)van der Veen,Habing, Geballe]{vanderVeen89}
 van der Veen, W.E.C.J., Habing, H. J.,
 Geballe, T. R.
 1989, A\&A 226, 108

\bibitem[Van de Steene et al.(2000)Van de Steene, van Hoof, Wood]{VandeSteene00}
 Van de Steene, G.C., van Hoof, P.A.M.,
 Wood, P.R.
 2000, A\&A 362, 984


\bibitem[Waters et al.(1996)]{Waters96}
 Waters, L.B.F.M., Molster, F.J., de Jong, T., et al.
 1996, A\&A 315, L361

\bibitem[Waters et al.(1998)]{Waters98}
 Waters, L.B.F.M., Cami, J., de Jong, T., et al.
 1998, Nature 391, 868

\bibitem[Waelkens et al.(1996)]{Waelkens96}
 Waelkens, C., Van Winckel, H., Waters, L.B.F.M., Bakker, E.J.
 1996, A\&A 314, L17

\bibitem[Zijlstra et al.(1991)]{Zijlstra91}
 Zijlstra A.A.,  Gaylard M.J., te Lintel Hekkert, P., Menzies, J.W.,
 Nyman, L.-A., Schwarz, H. E., 1991, A\&A, 243, L9




\end{thebibliography}

\end{document}